\documentstyle[preprint,aps,amsfonts]{revtex}

\begin{document}
\draft
\title{{\rm \normalsize\hfill Imperial/TP/95-96/63\\
 \hfill DAMTP R96/44\\}
	Information-entropy and the space of decoherence
                  functions in generalised quantum theory}

\author{C.J.~Isham\thanks{email: c.isham@ic.ac.uk}}
\address{Theoretical Physics Group,
		Blackett Laboratory\\
		Imperial College of Science, Technology \& Medicine\\
		South Kensington, London  SW7 2BZ, U.K.}
\author{N.~Linden\thanks{email: n.linden@newton.cam.ac.uk}}
\address{D.A.M.T.P.\\
        University of Cambridge\\
        Cambridge CB3 9EW, U.K.
}

\date{November, 1996}
\maketitle

\begin{abstract}
In standard quantum theory, the ideas of
information-entropy and of pure states are closely linked.  States
are represented by density matrices $\rho$ on a Hilbert space and
the information-entropy $-\mbox{tr}(\rho\log\rho)$ is minimised on
pure states (pure states are the vertices of the boundary of the
convex set of states).  The space of decoherence functions in the
consistent histories approach to generalised quantum theory is
also a convex set.  However, by showing that every decoherence
function can be written as a convex combination of two other
decoherence functions we demonstrate that there are no `pure'
decoherence functions.

The main content of the paper is a new notion of
information-entropy in generalised quantum mechanics which is applicable in
contexts in which there is no {\em a priori} notion of time. Information-entropy
 is defined first on consistent sets and
then we show that it decreases upon refinement of the consistent
set.  This information-entropy suggests an intrinsic way of giving a
consistent set selection criterion.
\end{abstract}

\section{INTRODUCTION}
A particularly attractive feature of the consistent
histories programme, as developed by Gell-Mann and Hartle
\cite{GH90a,GH90b,GH90c,Har91a,Har91b,GH92,Har93a} following
pioneering work by Griffiths \cite{Gri84} and Omn\`es
\cite{Omn88a,Omn88b,Omn88c,Omn89,Omn90,Omn92} is that it offers a 
framework for quantum theory in which time potentially plays a
subsidiary r\^ole\footnote{For more recent developments
in the consistent histories programme by these authors see, for example,
\cite{GH95,Gri95,Omn94}.}.  The central idea of the scheme is that under
certain {\em consistency\/} conditions it is possible to assign
probabilities to generalised histories of a system.  In normal
quantum theory such histories are represented by time-ordered
strings of propositions; however the scheme allows for much more
general histories in which there is no {\em a priori\/} notion of
time ordering.  These generalised histories are expected to play a
key r\^ole in application of the formalism to quantum gravity.

	In the generalised version of the history scheme that we have
developed \cite{Isham94a,IL94a,IL94b} the central mathematical
ingredients are a set of histories ${\cal UP}$ (or, more accurately, the
set of {\em propositions\/} about histories) and an associated set of
decoherence functions $\cal D$, with the pair $({\cal UP},{\cal D})$
being regarded as the analogue in the history theory of the pair
$({\cal L},{\cal S})$ in standard quantum theory where $\cal L$ is
the lattice of propositions and $\cal S$ is the space of states on
$\cal L$.

	In this paper we address two related issues: we investigate the
structure of the convex set of decoherence functions and we
suggest a new definition of information-entropy for decoherence
functions.  The analogues of these ideas in standard quantum
theory are simply related by the fact that the information-entropy
$I_{single-time} = -\mbox{tr}(\rho\log\rho)$ is minimised on the
vertices of the boundary of the convex set of density matrices
$\rho$; these vertices are the `pure' density matrices corresponding
to pure states in the Hilbert space.  As we shall show, although
convex, the space of decoherence functions has a very different
structure and there are no `pure' decoherence functions.  For this
and other reasons we need a rather different approach to the notion
of information-entropy in generalised quantum theory.

	Several other authors \cite{Har95,Halliwell93,Kent96,McElwaine96} have considered
aspects of information theory in the context of the consistent
histories approach.  In particular, in a very interesting paper that
partly motivated our work, Hartle \cite{Har95} proposed a definition
of information that we describe in section 3.  We feel however that
our alternative definition has certain advantages over that given in
\cite{Har95}: in particular, it is more straightforward.

\section{THE CONVEX SET OF DECOHERENCE FUNCTIONS}

\subsection{`Pure' decoherence functions}
In \cite{Isham94a,IL94a} we described how the space ${\cal UP}$ encodes
the generalised quantum temporal logic of the propositions.  As
explained in \cite{Isham94a,IL94a}, there are compelling reasons for
postulating that the natural mathematical structure on ${\cal UP}$ is that
of an {\em orthoalgebra\/} \cite{FGR92}, with the three orthoalgebra
operations $\oplus$, $\neg$, and $<$ corresponding respectively to
the disjoint sum, negation, and coarse-graining operations invoked by
Gell-Mann and Hartle.  One example of an orthoalgebra is the lattice
of projection operators on a Hibert space.  In this case, the
operation $\oplus$ is defined on disjoint pairs of projectors $P,Q$
with $P\oplus Q = P\vee Q$ where, as usual, $P\vee Q$ denotes the
projector onto the linear span of the subspaces onto which $P$ and
$Q$ project.  In the example of a lattice (which is a special type of
orthoalgebra), $\vee$ is defined on all
projectors, not only on pairs that are disjoint.

	Throughout this paper we shall be dealing with the case where
the orthoalgebra of propositions is the space of projectors on a
Hilbert space ${\cal V}$ which, for the sake of simplicity, we shall take
to be finite dimensional.  This Hilbert space may arise from having
propositions at $n$ time points, in which case ${\cal V} = \otimes^n {\cal H}$
(see below), but it need not do so.  A crucial ingredient in our
construction of the information-entropy will be the {\em
dimension\/} of a proposition, defined to be the dimension of the
projector that represents the proposition on ${\cal V}$.

	The properties of the decoherence function $d:
{\cal UP}\times{\cal UP}\rightarrow
{\Bbb C}$ are  
\begin{enumerate}
    \item {\em Hermiticity\/}: $d(\alpha,\beta)=d(\beta,\alpha)^*$ for all
          $\alpha,\beta\in{\cal UP}$. 
    \item {\em Positivity\/}: $d(\alpha,\alpha)\ge0$ for all $\alpha\in{\cal UP}$.
    \item {\em Additivity\/}: if $\alpha$ and $\beta$ are disjoint then, for
          all $\gamma$, $d(\alpha\oplus\beta,\gamma)=d(\alpha,\gamma)+d(\beta,\gamma)$. 
    \item {\em Normalisation\/}: $d(1,1)=1$.  
\end{enumerate} 

	One important motivation for our framework is the fact that
discrete-time histories in quantum theory can indeed be given the
structure of an orthoalgebra. The key idea is that an $n$-time,
homogeneous history proposition
$(\alpha_{t_1},\alpha_{t_2},\ldots,\alpha_{t_n})$ can be associated with the
operator $\alpha_{t_1}\otimes\alpha_{t_2}\otimes\cdots\otimes\alpha_{t_n}$ which
is a genuine {\em projection\/} operator on the $n$-fold tensor
product ${\cal H}_{t_1}\otimes{\cal H}_{t_2}\otimes\cdots\otimes{\cal H}_{t_n}$ of
$n$-copies of the Hilbert-space ${\cal H}$ on which the canonical theory
is defined \cite{Isham94a,IL94a}.

	It may be noted that if $d_1$ and $d_2$ are decoherence
functions then so is
\begin{equation}
	d_{(\lambda)} := \lambda d_1 + (1-\lambda)d_2
\end{equation}
where $\lambda$ is a real constant $0\leq \lambda \leq 1$. Thus the space
of decoherence functions is a convex set.

	What we have said so far looks very similar to the situation in
standard single-time quantum theory where---by the use of Gleason's
theorem---a state may be characterised by a positive self-adjoint
operator with trace $1$ ({\em i.e.,\ } a density matrix) on the Hilbert space.
The probability, $\hbox{\rm Prob}(P;\rho)$, that the proposition represented by the projection
operator $P$ is true if the system is in the state $\rho$ and a
suitable measurement is made is
\begin{equation} 
	\hbox{\rm Prob}(P;\rho)=\mbox{tr}(P\rho).
\label{rhoprob}
\end{equation}

The space of density matrices is also convex in the sense that
\begin{equation}
	\rho_{(\lambda)} := \lambda\rho_1+ (1-\lambda)\rho_2 			\label{rho}
\end{equation}
is a state if $\rho_1$ and $\rho_2$ are states and $0\leq\lambda\leq 1$.  In
standard single-time quantum theory a state is said to be {\em
pure\/} if it cannot be written in the form (\ref{rho}) with $\rho_1
\neq \rho_2$.  Pure states play an important r\^ole since, in this case, the
probabilities (\ref{rhoprob}) cannot be interpreted as arising from a
stochastic mixture.

	Since  a state $\rho$ is a positive self-adjoint operator, the
spectral theorem shows that it can be written as
\begin{equation}
	\rho = \sum_i r_i P_i
\end{equation}
where $0\leq r_i \leq 1$ are the eigenvalues of $\rho$ and $P_i$ are
the projectors onto the associated eigenspaces.  This shows that
unless all the $r_i$ are zero except one, the state is certainly
impure.  Furthermore, it can be shown \cite{Vara68} that pure states are
of the form
\begin{equation}
	\rho = P,
\end{equation}
where $P$ is a projection operator onto a one-dimensional subspace.

	In generalised history quantum theory, although the set of
decoherence functions is convex, in other respects the situation is quite different from that in standard quantum theory.
One may attempt to define a pure decoherence function $d$ as one that
cannot be written in the form
\begin{equation}
	d = \lambda d_1 + (1-\lambda)d_2
\end{equation}
with $d_1\neq d_2$. However we now show that there are no such
decoherence functions.

	Firstly, let us recall \cite{IL94b} that we have characterised
all decoherence functions in the case that ${\cal UP}$ is the lattice of
projectors on a finite dimensional Hilbert space as follows 
\footnote{Generalisations of this result have been given in 
\cite{Wright95,Rudolph96}.}.
 
{\sl Decoherence functions are in one-to-one correspondence with
`decoherence operators' $X$ on ${\cal V}\otimes{\cal V}$ according to the rule
\begin{equation}
	 d(\alpha,\beta) = tr_{{\cal V}\otimes{\cal V}}(\alpha\otimes\beta X)
\end{equation}
where the decoherence operator $X$ satisfies
\begin{enumerate}
\item   $MXM = X^\dagger$ where $M(u\otimes v) := v \otimes u$; 
\item  $\mbox{tr}_{{\cal V}\otimes{\cal V}}(\alpha\otimes\alpha X) \ge 0$; 
\item  $\mbox{tr}_{{\cal V}\otimes{\cal V}}(X) = 1$.         
\end{enumerate}
}
It should be noted that $X$ need not be a positive operator.  Indeed, in
\cite{IL94a} we found examples of decoherence functions in standard quantum
theory where $d(\alpha,\alpha)>d(\beta,\beta)$ for two histories $\alpha$ and $\beta$
for which $\alpha\leq\beta$, and we also found decoherence functions and
histories $\gamma$ for which $d(\gamma,\gamma)>1$.

	Now consider an operator on ${\cal V}\otimes{\cal V}$ of the following form
\begin{equation}
	Y= i(s_1\otimes s_2 -s_2\otimes s_1)
\end{equation}
for any self-adjoint operators $s_1,\ s_2$ on ${\cal V}$.  It may be seen
that $Y$ satisfies
\begin{enumerate}
\item  $MYM = Y^\dagger$ where $M$ is the interchange operator given
above;
\item  $\mbox{tr}_{{\cal V}\otimes{\cal V}}(\alpha\otimes\alpha Y) = 0 \quad \forall \alpha$; in 
particular $\mbox{tr}_{{\cal V}\otimes{\cal V}}(Y) =
\mbox{tr}_{{\cal V}\otimes{\cal V}}(1\otimes 1\ Y) = 0$.
\end{enumerate}
Given any decoherence operator $X_d$ associated with a decoherence
function $d$, let us define two new operators $X_d^+ $ and $ X_d^- $
by
\begin{equation}
	X_d^+ = X_d + Y \quad {\rm and}\quad X_d^- = X_d - Y.
\end{equation}
Then $X_d^+ $ and $ X_d^- $ are also decoherence operators, as may easily
be checked.

	Now consider the identity
\begin{equation}
	X_d \equiv  \frac{1}{2} (X_d + Y) + \frac{1}{2}(X_d - Y) = 
		\frac{1}{2} X_d^+ +\frac{1}{2} X_d^-.
\end{equation}
It is clear that if $d^+$ and $d^-$ denote the decoherence functions
that are associated with the decoherence operators $X^+$ and $X^-$
respectively then
\begin{equation}
	d = \frac{1}{2} d^+ +\frac{1}{2} d^- 
\end{equation}
and hence $d$ is impure.  Thus there are no `pure' decoherence
functions\footnote{Other aspects of the structure of the space of 
decoherence functions have been considered in \cite{Schr96a}.}.

\subsection{Pure decoherence functions with respect to a window}
Whilst there are no pure decoherence functions in general, it is
possible to discuss a notion of purity of a decoherence function 
in the context of a fixed consistent set. In general we shall refer
to an exclusive and exhaustive set of propositions ({\em i.e.} a resolution of 
the identity in the orthoalgebra ${\cal UP}$) as 
a {\em window\/} $W =\{\alpha_i\}$ \footnote{Note that in \cite{Isham96} the word window
is used to describe the Boolean algebra generated by this set of propositions, 
rather than the set of propositions itself.}.

	Firstly, we define two decoherence functions, $d_1$ and $d_2$ to
be $W$-{\em equivalent} if
\begin{enumerate}
\item $W$ is a consistent set with respect to both $d_1$ and $d_2$ 
 ({\em i.e.,\ } $d_1(\alpha_i,\alpha_j) = d_2(\alpha_i,\alpha_j) = 0$ for all $\alpha_i,\alpha_j\in
W$ with $\alpha_i\neq\alpha_j$); and

\item $d_1(\alpha_i,\alpha_i) = d_2(\alpha_i,\alpha_i)$ for all $\alpha_i\in W$.
\end{enumerate}
It may readily be checked that this is indeed an equivalence
relation on  the space $\cal D$ of all decoherence functions.

	Each equivalence class of $W$-equivalent decoherence functions
may be represented by the member $d_{\tilde X}$ whose decoherence
operator has the `canonical form' (this decoherence function has
been useful in other contexts, see \cite{Schr95})
\begin{equation}
	\tilde X = \sum_{i=1}^{n} {d(\alpha_i,\alpha_i)\over (\dim \alpha_i)^2} 
		\alpha_i\otimes\alpha_i.					\label{tildex} 
\end{equation}
We shall shortly be making use of the fact that $\tilde X$ is a positive
self-adjoint operator on ${\cal V}\otimes{\cal V}$.

	Let us denote the space of $W$-equivalence classes by ${\cal
D}_W$.  Then  $\delta\in{\cal D}_W$ is said to be
$W$-{\em pure} if it cannot be written in the form
\begin{equation}
	\delta = \lambda\delta_1 + (1-\lambda)\delta_2 \label{delta_mixed}
\end{equation}
 with $\delta_1 \neq \delta_2$; if $\delta$ can be written in the form
(\ref{delta_mixed}) with $\delta_1 \neq \delta_2$ with $0\leq\lambda\leq 1,\ \lambda \in {\Bbb R}$, we say it is $W$-{\em
impure}. Note that the sum of  equivalence classes on the right
hand side of (\ref{delta_mixed}) is well-defined since if $d_1\equiv_W
d_1'$ and $d_2\equiv_W d_2'$ then $\lambda d_1+(1-\lambda)d_2\equiv_W \lambda
d_1'+(1-\lambda)d_2'$, where $\equiv_W$ means $W$-equivalent. 

	Clearly any $\delta$ that can be represented by a decoherence
operator $\tilde X$ of the form (\ref{tildex}) with more than one
non-zero $d(\alpha,\alpha)$ is impure.  On the other hand, consider a
decoherence operator 
\begin{equation}
	\tilde X :=  {1\over(\dim \alpha)^2} \alpha\otimes\alpha	\label{purex} 
\end{equation}
where $\alpha$ is one of the members of $W$ and suppose that its
associated $W$-equivalence class of decoherence functions $\delta$ can be
decomposed in the form (\ref{delta_mixed}).  Now $\delta_1(\beta,\beta)\geq 0$
and $\delta_2(\beta,\beta)\geq 0$ for all $\beta$.  Also, since $\delta (\beta,\beta)=0$
for all $\beta\in W$ such that $\beta\neq\alpha$, we have
\begin{equation}
	\delta (\beta,\beta) = 0 = \lambda\delta_1(\beta,\beta) + (1-\lambda)\delta_2(\beta,\beta) 
		\quad\forall\beta\in W\mbox{ such that }\beta \neq \alpha
\end{equation}
and therefore
\begin{equation}
	0 = \delta_1(\beta,\beta)=\delta_2(\beta,\beta) \quad \forall 
			\beta\in W\mbox{ such that }\beta\neq\alpha.
\end{equation}
Also $\delta_1(1,1) = 1$ and $\delta_2(1,1)=1$, which implies
$\delta_1(\sum_{\alpha_i\in W} \alpha_i,\sum_{\alpha_j\in W} \alpha_j) = 1$ which in
turn implies $\delta_1(\alpha,\alpha) = 1$, and similarly $\delta_2(\alpha,\alpha) =1$.
Thus $\delta_1$ and $\delta_2$ are both equal to $\delta$ and hence $\delta$ is pure.

\section{INFORMATION-ENTROPY}
We turn now to the question of defining the information-entropy in the context 
of a window and for a given decoherence function.  In standard single-time
quantum theory the information-entropy is given by
\begin{equation}
	I_{s-t} = -\mbox{tr}(\rho\log\rho) 				\label{In_single}
\end{equation}
where $\rho$ is the density matrix, and, as mentioned in the Introduction,
$I_{s-t}$ is minimised on pure states.

	What we seek is a notion of information-entropy that can be used
in generalised history quantum theory.  In particular, the
definition should be applicable in principle to systems in which the
concept of time is not fundamental and may emerge only in some
coarse-grained way.  Furthermore, even if the system has a standard notion of
time, the information-entropy---which encodes the number of bits
required to describe the system---may not necessarily all reside in
the initial state.  The description of a system in this generalised
type of quantum theory is given entirely in terms of the set of
propositions and the values of the decoherence function, so we must
construct our measure of information-entropy solely from these.

	Firstly, however, we point out that since---as explained
above---the decoherence function can be described in terms of a
decoherence operator $X$, the most na{\"\i}ve approach (without
physical motivation) might be to try to construct a measure of
information-entropy for a decoherence function from $X$. The
simplest analogue of (\ref{In_single}) is 
\begin{equation}
	I_d = -\mbox{tr}(X\log X) 
\end{equation} 
but this is not defined in general since $X$ is neither self-adjoint
nor positive.

	However, focussing on the probability distributions derived from
$d$ does, in fact, offer a way of defining information-entropy.  To
see this consider a general probability distribution with $M$ events
$\{e_i\}_{i=1}^M$ with probabilities $\{\mbox{Prob}(e_i)\}_{i=1}^M$.  The
usual measure of the information-entropy of this distribution is
\begin{equation}
-\sum_{i=1}^M \mbox{Prob}(e_i)\log\mbox{Prob}(e_i).
\end{equation}
On the other hand, a given decoherence function produces not one but
many probability distributions, namely one for each consistent
window. A possible start, therefore, might be to define the
information-entropy in the context of a window $W=\{\alpha_i\}$ as
\begin{eqnarray}
I_W^{trial} &=& -\sum_i\mbox{Prob}(\alpha_i)\log\mbox{Prob}(\alpha_i)\nonumber\\
            &=& -\sum_i d(\alpha_i,\alpha_i) \log d(\alpha_i,\alpha_i)\label{I_W_trial}
\end{eqnarray}
which is indeed now well-defined. However, as noted by Hartle
\cite{Har95}, $I_W^{trial}$ does not have the appropriate properties
with respect to refinement of the consistent set.  In particular, we
require that if the consistent set is refined---corresponding to
having finer-grained propositions---the information-entropy should
decrease or stay the same.  However (\ref{I_W_trial}) does not have
this property---indeed, the most coarse grained set $W=\{0,1\}$
(where $1$ is the projector onto the whole Hilbert space) is always
consistent and has $d(1,1) = 1$ and $d(0,0)=0$, so that
$I_W^{trial}=0$.  As Hartle \cite{Har95} puts it, there is no
penalty for asking stupid questions.

	In his very interesting paper \cite{Har95}, Hartle considered
this problem and proposed the following definition of `space-time
information-entropy'. First choose a measure of the
missing information $S(d)$ in the decoherence function $d$;  for
example, one could choose a standard class ${\cal C}_{\rm
stand}$ of consistent sets and then define $S(d)$ as
\begin{equation}
	S(d) := \min_{W \in {\cal C}_{\rm stand} } \left[
		-\sum_\alpha d(\alpha,\alpha) \log\ d(\alpha,\alpha)\right],
\end{equation}
where $W$ is varied over all consistent sets of histories in
the standard class.  Hartle suggests that ${\cal C}_{\rm stand}$
might be chosen to be the class of {\em finest\/} grained histories
that decohere.

	Having chosen $S(d)$ the next step is to use the Jaynes
construction \cite{Jaynes} to define the missing information in a
general set of decoherent histories $W$.  The missing
information is the maximum of the information content of decoherence
functions which reproduce the decoherence and probabilities of the
set $W$:
\begin{equation}
	S(W,d) := \max_{\tilde d } \left[S(\tilde d)
			\right]_{\tilde d(\alpha,\alpha)=d(\alpha,\alpha)},
\end{equation}
where the maximum is taken over all decoherence functions $\tilde d$
that reproduce the decoherence function for the set of histories
$W$. Finally, the missing information in any class ${\cal C}$
of decoherent sets of histories is  defined as 
\begin{equation}
S({\cal C},d) = \min_{W \in {\cal C} } \left[S(W,d)\right].
\end{equation}

	We feel that the definition of information-entropy that we shall
now develop has a number of potential advantages over that given by Hartle.
In particular (i) it is fairly simple; (ii) it does not need the use of
maximum entropy ideas; (iii) it does not require the choice a
standard class of consistent sets.  We shall also show in the
next section, when calculated in the case of standard quantum
theory (for consistent sets of homogeneous histories) the information-entropy
for the decoherence function is found
to be equal to $-\mbox{tr}(\rho\log\rho)$, up to normalisation,  where $\rho$ is the
initial density matrix. 

	As a first step towards finding this new definition of
information-entropy consider any window $W =
\{\alpha_i\}_{i=1}^n$ that is consistent with respect to a given
decoherence function $d$.  Then, as explained above, the canonical
decoherence operator $\tilde X_{d,W}$ that reproduces the values of
$d(\alpha_i,\alpha_j)$ of $d$ in the window $W$ is
\begin{equation}
\tilde X_{d,W} = 
	\sum_{i=1}^n {d(\alpha_i,\alpha_i)\over (\dim \alpha_i)^2} \alpha_i\otimes\alpha_i.
\end{equation} 
The crucial observation is that, unlike a general decoherence
operator, $\tilde X_{d,W}$ {\em is\/} a positive, self-adjoint
operator on ${\cal V}\otimes{\cal V}$, and so one can define the logarithm of
$\tilde X_{d,W}$ and thereby form
\begin{equation}
 -\mbox{tr}(\tilde X_{d,W}\log\tilde X_{d,W})
= -\sum_{i=1}^n d(\alpha_i,\alpha_i)\log \left({d(\alpha_i,\alpha_i)\over (\dim\alpha_i)^2}\right).
\end{equation}
This motivates the following definition of the 
information-entropy for this decoherence function and window:
\begin{equation}
	\hat I_{d,W}:= -\sum_{i=1}^n d(\alpha_i,\alpha_i)
		\log{d(\alpha_i,\alpha_i)\over (\dim\alpha_i)^2}. 			\label{hatIdW}
\end{equation}

While this function has many of the propeties that are desired of a measure
of information, as we shall show below, there are persuasive arguments 
\footnote{We are extremely grateful to J Hartle and A Kent for reading an earlier
draft of this paper and bringing these issues to our attention.}
for renormalising this function and to define our measure of information as

\begin{eqnarray}
        I_{d,W} &:=& \hat I_{d,W} -\hat I_{d,\{1,0\}}\nonumber\\
&=& 
-\left(\sum_{i=1}^n d(\alpha_i,\alpha_i)
                \log{d(\alpha_i,\alpha_i)\over (\dim\alpha_i)^2}\right)-\log\dim{\cal V}^2\nonumber\\
&=& -\sum_{i=1}^n d(\alpha_i,\alpha_i)
                \log{d(\alpha_i,\alpha_i)\over (\dim\alpha_i/\dim{\cal V})^2},
        \label{IdW}
\end{eqnarray}
where $\hat I_{d,\{1,0\}}$ is the value of $\hat I_{d,W}$ for the (coarsest)
 window $\{1,0\}$, and ${\cal V}$ is the Hilbert space on which the history
propositions are defined.

The function 
\begin{equation}
{\dim\alpha_i\over \dim{\cal V}}
\end{equation}
is the {\em relative} dimension of the projector $\alpha_i$.  
Before we go further to describe properties of this measure
of information-entropy, let us describe the reasons for this use of relative
dimension rather than absolute dimension of a proposition.  The difference is
clearly only important if one envisages comparing information-entropy in situations
where ${\cal V}$ changes.  An important case in point is the history version
of $n$-time quantum mechanics.  In this case, ${\cal V}=\otimes^n {\cal H}$,
where ${\cal V}=\otimes^n {\cal H}$ is the Hilbert space of the 
canonical theory.  One uses $n$-times to model a situation in 
which `nothing happens' in the intermediate times.  

Consider a consistent set 
of $n$-time histories
\begin{equation}
\{\alpha_i\} = \{P^i_1,P^i_2...P^i_{t_r},P^i_{t_{r+1}},...P^i_n\}.
\end{equation}
Now imagine inserting an additional time, between ${t_r}$ and $t_{r+1}$, say, 
but use the unit projector at this time.  The use of relative dimension ensures
that the information-entropy does not change when one does this 
trivial extension to the consistent set, since the dimension of the history
\begin{equation}
\alpha_i = (P^i_1,P^i_2...P^i_{t_r},1,P^i_{t_{r+1}},...P^i_n)
\label{extended}\end{equation}
is $\dim{\cal H}$ times that of
\begin{equation}
\alpha_i = (P^i_1,P^i_2...P^i_{t_r},P^i_{t_{r+1}},...P^i_n)
\label{unextended}
\end{equation}
however the dimension of ${\cal V}$ in (\ref{extended}) is $\dim{\cal H}$ times that of (\ref{unextended}).

An additional aspect of 
the use of relative dimension is that it may help in extending our work to
infinite dimensions, since it may be possible to use von Neumann's theory
of dimension functions of type $\hbox{\rm II}_1$ algebras of projectors\cite{MvN36}.

Returning now to consideration of general properties of (\ref{IdW}), 
we note that the definition we have given is close to the simple form (\ref{I_W_trial}), 
however the
extra factor $(\dim\alpha_i)^{-2}$ is the crucial ingredient that
results in the thus-defined information-entropy being either
constant or decreasing when the window is refined. To see this,
consider two consistent windows $W_1=\{\alpha,\alpha_1,\alpha_2\ldots,\alpha_n\}$
and $W_2=\{\beta,\gamma,\alpha_1,\alpha_2\ldots,\alpha_n \}$ where $W_2$ is a
refinement of $W_1$ in the sense that one of the projection
operators in $W_1$, namely $\alpha$, is split into two with
$\alpha=\beta\oplus\gamma$.  Thus
\begin{eqnarray}
I_{d,W_1}-I_{d,W_2} &= - d(\alpha,\alpha) \log \left({d(\alpha,\alpha)\over(\dim\alpha)^2}\right)
                       +  
d(\beta,\beta)\log\left({d(\beta,\beta)\over(\dim\beta)^2}\right)\nonumber\\ 
&\quad + d(\gamma,\gamma)\log\left({d(\gamma,\gamma)\over(\dim\gamma)^2}\right).
\end{eqnarray}
For simplicity of notation it will be convenient to define the ratios
 \begin{equation}
	a:={d(\gamma,\gamma) \over d(\beta,\beta)} \mbox{ and } 
		b:={\dim(\gamma)\over\dim(\beta)}
\end{equation}
and, without loss of generality, we can take $0\leq a<\infty$ and
$1\leq b<\infty$. Now $d(\alpha,\alpha) = d(\beta,\beta) + d(\gamma,\gamma)$ and $\dim(\alpha)
= \dim(\beta) + \dim(\gamma)$, and hence 
\begin{equation}
	I_{d,W_1}-I_{d,W_2} = d(\beta,\beta)\left( 
		a\log\left({a\over b^2}\right)
		- (1+a)\log\left({(1+a)\over(1+b)^2}\right)\right).
\end{equation}
It is not too difficult to prove the inequality
\begin{equation}
 a\log\left({a\over b^2}\right)
- (1+a)\log\left({(1+a)\over(1+b)^2}\right) 
    \geq 0\quad{\rm for}\quad 0\leq a<\infty \ {\rm and}\  1\leq b<\infty,
\end{equation}
which implies that $I_{d,W_1}$ decreases with respect to this special
type of refinement.  However, {\em any} refinement of $W_1$ can be
reached in a step-wise fashion by repeated refining of one
projection operator into two, and hence $I_{d,W}$ decreases
under any refinement.  We note in particular that with this
definition of $I_{d,W}$ the consistent set $W=\{1,0\}$ has
information-entropy $0$ and that this is the maximum possible value of the 
function $I_{d,W}$; the minimum possible value of 
\begin{equation}
 \hat I_{d,W}:= -\sum_{i=1}^n d(\alpha_i,\alpha_i)
                \log{d(\alpha_i,\alpha_i)\over (\dim\alpha_i)^2}.
\end{equation}
for any $d$ and $W$ is zero (which occurs if there is a consistent set
all of whose projectors are one-dimensional and all of whose probabilities,
bar one, are zero)
so that the minimum value of $ I_{d,W}$ is $-2\log\dim{\cal V}$.

	At this stage we might proceed in several different ways.  One
possibility is to leave the information-entropy defined in
this `localised' form $I_{d,W}$ in which the context $W$ 
appears explicitly. This procedure would be rather natural within
the topos-theoretic interpretation of the consistent histories
formalism that was introduced recently by one of us \cite{Isham96}.
In this case it is appropriate to define $I_{d,W_0}$ for
{\em any\/} window $W_0$ ({\em i.e.,\ } any set of exclusive and exhaustive
histories, not necessarily one that is $d$-consistent) as
\begin{equation}
	I_{d,W_0}:=\min_{W\geq W_0}I_{d,W}
\end{equation}
where the minimisation is taken over all coarse-grainings $W$ of
$W_0$ that {\em are\/} $d$-consistent. We note that $W_0\mapsto
I_{d,W_0}$ is an order-preserving map from the partially-ordered set
of windows to the ordered set of real numbers---an essentially
`functorial' property in the language of \cite{Isham96}.

	A second possibility is to define {\em the} information-entropy
of the decoherence function $d$ as the minimum over all consistent
sets of $I_{d,W}$, {\em i.e.,\ }
\begin{equation}
	I_{d}:= \min_W I_{d,W}.					\label{Id}
\end{equation} 
As will become clear from the examples below the consistent set (or sets) which minimise
$I_{d,W}$ seem to be naturally associated with the decoherence operator $X$.

	Before proceeding to illustrate these ideas with examples drawn
from standard $n$-time quantum theory it is worth emphasising
that this definition of $I_{d}$ is non-trivial; in particular, it is
not independent of $d$ (as, {\em a priori\/}, it might have been). To
demonstrate this point consider the decoherence function associated
with the following decoherence operator on the space ${\cal V}\otimes {\cal V}$
with ${\cal V} =
{\Bbb C}^2$:
\begin{equation}
	X_1 := \frac{1}{2}\left(\alpha\otimes\beta + \beta\otimes\alpha \right)
\end{equation}
where
\begin{equation}
\alpha =  \left(\begin{array}{ll}  
					1 & 0  \\
					0 & 0\\ 
					\end{array}\right),
\quad \beta =  \left(\begin{array}{ll}  
					0 & 0  \\
					0 & 1\\ 
					\end{array}\right).
\end{equation}
If $P$ is the most general one-dimensional projection operator
\begin{equation}
P=  \left(\begin{array}{lc}  
					a & b  \\
					b^* & 1-a\\ 
					\end{array}\right)
\end{equation}
with $a\in{\Bbb R},\ 0\leq a\leq 1$ and $b\in{\Bbb C},\ |b|^2 = a(1-a)$
one may easily calculate that 
\begin{equation}
d(P,1-P) = \frac{1}{2} \left( a^2 + (1-a)^2\right) 
\end{equation}
so that there are no one-dimensional consistent sets and the only
consistent set is $\{0,1\}$ which has $I_{d_1} = 0$. On the
other hand, the decoherence function $d_2$ associated with the
decoherence operator
\begin{equation}
X_2 = \alpha \otimes \alpha
\end{equation}
on the same space has $I_{d_2} = -2\log 2$ (since in this case the set  $\{\alpha,
\beta\}$ is consistent and 
 has $I_{d_2} = -2\log 2$).

\section{EXAMPLES}
\subsection{The history version of standard quantum theory}
The definition we have given for $I_d$ is in terms of consistent
sets and their associated probability distributions and it is
interesting to see how it reduces up to normalisation to the familiar
information-entropy
\begin{equation}
	I_{s-t} = -\mbox{tr}(\rho\log\rho)
\end{equation}
of standard quantum theory. In this case the histories
are simply projectors at one time point and the value of the
decoherence function is
\begin{equation}
	d(P,Q) = \mbox{tr}(P\rho Q).
\end{equation}
Firstly it should be noted that all exhaustive and exclusive sets are
consistent since if $P_1$ and $P_2$ are two orthogonal projectors
then
\begin{equation}
d(P_1,P_2) = \mbox{tr}(P_1\rho P_2) = \mbox{tr}(\rho P_2P_1) = 0.
\end{equation}
Secondly since---as shown in the previous section---the
information-entropy is constant or decreases upon refinement of the
window, it suffices to consider windows in which all the projectors
are one-dimensional.  Thus the information-entropy of the
decoherence function is the minimum over all one-dimensional
resolutions of the identity $W=\{P_i\}_{i=1}^N$ of $I_{d,W}$ (where
$N$ is the dimension of the Hilbert space ${\cal V}$; of course, in this
case, ${\cal V}$ is just the canonical Hilbert space ${\cal H}$) 
\begin{eqnarray}
	I_{d,W} &=& -\sum_i d (P_i, P_i) \log d(P_i, P_i) -2\log N
		\nonumber\\ 
		&=& -\sum_i\mbox{tr}(P_i\rho P_i)\log\mbox{tr}(P_i\rho P_i)-2\log N\nonumber\\ 
		&=& -\sum_i\mbox{tr}(\rho P_i)\log\mbox{tr}(\rho P_i)-2\log N.
\end{eqnarray}
This expression for $I_{d,W}$ is independent of the basis in which the traces
are evaluated and it is convenient to evaluate it in the basis in which the
density matrix, $\rho$, is diagonal:
 \begin{equation} 
I_{d,W}  = -\sum_i   \left(\sum_j(P_i)_{jj} r_j\right) \log  
\left(\sum_j(P_i)_{jj} r_j\right)-2 \log N
\end{equation}
where the $(P_i)_{jj}$ are the diagonal elements of $P_i$ in this basis and
$r_j$ are the (possibly repeated) eigenvalues of $\rho$.

Now the function $f(x) =-x\log x$ is a concave function and hence
satisfies the inequality
\begin{equation}
f\left( \sum_j \l_j x_j \right) \geq  \sum_j \l_j f(x_j) \label{Jensen}
\end{equation}
where the positive real numbers $\l_i$ satisfy $0\leq\l_i \leq 1$ and $\sum_i
\l_i = 1$ (this is essentially Jensen's inequality; see for example
\cite{CovTho91}). 

	 We shall now use this inequality to get a lower bound on $I_{d,W}$.  The
one-dimensional projectors have trace $1$, {\em i.e.,\ }
\begin{equation}
\sum_j (P_i)_{jj} = 1\ {\rm for\ each\ }i
\end{equation}
and therefore we can use (\ref{Jensen}) for each $i$ with $(P_i)_{jj}$
playing the r\^ole of $\l_j$, so that 
\begin{eqnarray} 
	I_{d,W} &=& -\sum_i\left(\sum_j(P_i)_{jj} r_j\right) \log
		\left(\sum_j(P_i)_{jj} r_j\right)-2\log N\nonumber\\
&\geq& - \sum_i \left( \sum_j (P_i)_{jj} \left( r_j\log r_j \right)
\right)-2\log{N}.
\end{eqnarray}
However, since $\sum_i P_i = 1$, in any basis we have
\begin{equation} 
	\sum_i (P_i)_{jj} = 1\ {\rm for\ each\ }j
\end{equation}
thus
\begin{eqnarray} 
	I_{d,W} &\geq& - \sum_i \left( \sum_j (P_i)_{jj} 
		\left( r_j\log r_j \right)\right)-2\log{N}				\nonumber\\
			&=& -\sum_j  \left( r_j\log r_j \right)-2\log{N}		\nonumber\\
			&=& -\mbox{tr}(\rho\log\rho)-2\log{N}.
\end{eqnarray}
We note that, if $\rho$ is non-degenerate, by choosing the $P_i$ to be
the spectral projections of $\rho$ we can indeed attain the bound.
Hence, if we define the information-entropy of the decoherence
function to be the minimum of $I_{d,W}$ over all $W$ we 
find
\begin{equation}
	I_{d}  = -\mbox{tr}(\rho\log\rho)-2\log{N}.
\end{equation}
If $\rho$ is degenerate, we should choose a resolution of the identity
by one-dimensional projectors obtained by replacing each
$n$-dimensional spectral projector $Q$ by any set of orthogonal
projectors which sum to $Q$; again one finds that $I_{d} =
-\mbox{tr}(\rho\log\rho)-2\log{N}$.

	Thus the definition of information-entropy that we have given
reduces, up to normalisation, to the usual one in the case of single-time quantum
theory.

\subsection{$n$-time quantum theory} 
We recall that in standard $n$-time quantum theory a history is a
time-ordered sequence of propositions about the system $\alpha=(P^{1}_{t_{1}}
,P^{2}_{t_{2}},\ldots,P^{n}_{t_{n}})$ with $t_1<t_2<\cdots<t_n$.  As we have
argued in \cite{Isham94a,IL94a}, this proposition should be
associated with the operator $(P^{1}_{t_{1}}\otimes
P^{2}_{t_{2}}\otimes\cdots\otimes P^{n}_{t_{n}})$ on the tensor product Hilbert
space ${\cal V} = {\cal H}_{t_1}\otimes{\cal H}_{t_2}\otimes\cdots\otimes{\cal
H}_{t_n}$ of $n$ copies of the Hilbert space ${\cal H}$ on which the canonical
theory is defined.  We have called histories such as
$(P^{1}_{t_{1}}\otimes P^{2}_{t_{2}}\otimes\cdots\otimes P^{n}_{t_{n}})$,
represented by a tensor product of operators on ${\cal V}$, {\em homogeneous};
there are, of course, many projectors on ${\cal V}$ which are not of this form.
In \cite{IL94b} we have shown how to construct the operator $X$ (an operator on
${\cal V}\otimes{\cal V}$) in this case so as to reproduce the standard
expression for the decoherence function, namely \begin{equation} 
	d(\alpha,\beta) = tr_{{\cal H}}(\tilde C_{\alpha}^\dagger
		\rho_{t_0} \tilde C_{\beta}) 
\end{equation} 
where
\begin{equation}
	\tilde C_{\alpha} = U(t_0,t_1)P^{1}_{t_{1}} U(t_1,t_2)P^{2}_{t_{2}}
		U(t_2,t_3)\ldots U(t_{n-1},t_n)P^{n}_{t_{n}} U(t_{n},t_0)
\end{equation}
and $U(t,t^\prime)=e^{-i(t-t^\prime)H}$ is the usual time-evolution
operator in the Hilbert space ${\cal H}$ of the canonical theory.

	We now show that the minimum value of the information-entropy
$I_{d,W}$ over all consistent sets of homogeneous histories for
standard quantum theory is
\begin{equation}
	-\mbox{tr}(\rho\log\rho)-2\log\dim{\cal V}.          \label{rhologrho}
\end{equation}
We suspect that this value is the minimum for {\em any} consistent
sets ({\em i.e.,\ } including inhomogeneous histories) but so far we have
only been able to prove this in certain examples (see below).

	Firstly we note that by taking the projection operators at each
time to be related to the spectral projectors of $\rho$ we can find a
consistent set that gives the value (\ref{rhologrho}) as the
information-entropy of that set.  More precisely, choose the
histories to be of the form
\begin{eqnarray}
\alpha &= (U(t_0,t_1)^{-1}P^{i}_{t_{1}} U(t_1,t_0)^{-1}, 
	U(t_0,t_2)^{-1}P^{j}_{t_{2}} U(t_2,t_0)^{-1},\ldots,	\nonumber\\
&\quad\ldots,U(t_0,t_n)^{-1}P^{n}_{t_{n}} U(t_n,t_0)^{-1})	\label{Uproj}
\end{eqnarray}
where $P_{i}$ is a one-dimensional projector onto the $i$th eigenspace of
$\rho$.  The unitary operators are needed to `undo' the time evolution so that
expressions for the probabilities become of the form
\begin{equation}
	\mbox{tr}(P_n\ldots P_jP_i\rho P_iP_j\ldots P_n).
\end{equation}
It is easy to see that all of the histories so defined will have
zero probability except those of the form $(P_i,P_i,\ldots,P_i)$,
$i=1,2,\ldots, N$ (where $N$ is the dimension of the Hilbert space)
which have probabilities
\begin{equation} 
	\mbox{tr}(P_i\ldots P_iP_i\rho P_iP_i\ldots P_i) = 
			\mbox{tr}(\rho P_i) = r_i
\end{equation}
where $r_i$ are the eigenvalues of $\rho$.  Thus the value of the
information-entropy in this window is
\begin{equation}
	I_{d,W} = -\sum_i^N r_i \log r_i -2\log\dim{\cal V}= -\mbox{tr}(\rho\log\rho)-2\log\dim{\cal V}.
\end{equation}

	In order show that the value of $I_{d,W}$ in any other window of
homogeneous projectors is greater than this it is helpful to note
the following. Let $\{Q^j\}_{j=1}^{N_1}$ be a resolution of
the identity by projectors in the Hilbert space ${\cal H}$, and let $K$ be
a positive self-adjoint operator.  Then
\begin{eqnarray}
 	-\sum_{j=1}^{N_1} \mbox{tr}(Q^j K Q^j) 
	\log\mbox{tr}\left({ (Q^j K Q^j) \over ( \dim Q^j)^2}\right)
		\qquad								
&=& -\sum_{j=1}^{N_1}\mbox{tr}(Q^j K ) 
\log\mbox{tr}\left({ ( Q^j K ) \over ( \dim Q^j)^2}\right)\nonumber\\
&\geq& - \mbox{tr}(K\log K),						\label{KlogK1}
\end{eqnarray}  
and also
\begin{equation}
	-\sum_{j=1}^{N_1}\mbox{tr}\left( (Q^j K Q^j) 
		\log \left({ (Q^j K Q^j) \over ( \dim Q^j)^2}\right)\right)
			\geq - \mbox{tr}(K\log K).				\label{KlogK2}
\end{equation} 

	The inequality (\ref{KlogK1}) is essentially the same as that proven in
the previous subsection (it should be noted that that proof did not
depend on the fact that $\rho$ had trace $1$).  The inequality
(\ref{KlogK2}) may most easily be seen by considering the left hand
side in a basis in which the $Q^j$'s are simultaneously diagonal.
Then the operator $Q^j K Q^j$ is of the block diagonal form
\begin{equation} 
Q^jK Q^j =\left(\begin{array}{lllll}(0)& \\ 
									&(0) \\ 
									& & (K^j) \\ 
									& & & (0) \\ 
									& & & & (0)
\end{array}\right)
\end{equation}
where $K^j$ is a $\dim Q^j\times \dim Q^j$ positive self-adjoint
matrix.  Clearly
\begin{eqnarray}
-\mbox{tr}_{N\times N}\left( Q^j K Q^j\log {Q^j K Q^j\over 
			( \dim Q^j)^2}\right)
	&=&-\mbox{tr}_{\dim Q^j\times \dim Q^j}\left( K^j\log {K^j\over 
		( \dim Q^j)^2}\right)							\nonumber\\
&\geq&-\mbox{tr}_{\dim Q^j\times \dim Q^j}\left( K^j\log {K^j}\right).
\end{eqnarray}

	We may now use these results and one from the previous section to
find an upper bound on the information-entropy for any consistent
window of homogeneous projectors. We have
\begin{eqnarray}
I_{d,W} &=& -\sum_{i_1,i_2,\ldots,i_{n-1},i_n} 
	\mbox{tr}(P_{i_n}P_{i_{n-1}}\ldots P_{i_2}P_{i_1}\rho
		P_{i_1}P_{i_2}\ldots P_{i_{n-1}}P_{i_n})\nonumber\\ 
& &\qquad\times \log\mbox{tr}\left(
{(P_{i_n}P_{i_{n-1}}\ldots P_{i_2}P_{i_1}\rho
P_{i_1}P_{i_2}\ldots P_{i_{n-1}}P_{i_n})\over  (\dim P_{i_1}\dim
P_{i_2}\ldots\dim P_{i_{n-1}}\dim P_{i_n})^2}\right)-2\log\dim{\cal V}\nonumber\\
 &=& 
-\sum_{i_1,\ldots,i_{n-1}}(\dim P_{i_1}\ldots\dim P_{i_{n-1}})^2 \nonumber\\
&& \quad\times\sum_{i_n} 
\mbox{tr}\left(P_{i_n}
\left[{P_{i_{n-1}}\ldots P_{i_1}\rho P_{i_1}\ldots P_{i_{n-1}}
\over 
(\dim P_{i_1}\ldots\dim P_{i_{n-1}})^2}\right]\right)\nonumber\\ 
& &\qquad\times \log \mbox{tr}\left(
{P_{i_n}\over (\dim P_{i_n})^2}
\left[{P_{i_{n-1}}\ldots P_{i_1}\rho P_{i_1}\ldots P_{i_{n-1}}
\over (\dim P_{i_1}\ldots\dim P_{i_{n-1}})^2}\right] \right)-2\log\dim{\cal V} 
\end{eqnarray}
and hence (\ref{KlogK1}) can be used to show that
\begin{eqnarray}
		\lefteqn{I_{d,W}} \nonumber\\
&\geq&
-\sum_{i_1,i_2,\ldots,i_{n-1}}(\dim P_{i_1}\ldots\dim P_{i_{n-1}})^2 \nonumber\\
& &\quad \times\quad\mbox{tr}\left(
\left[{
P_{i_{n-1}}\ldots P_{i_1}\rho P_{i_1}\ldots P_{i_{n-1}}
\over 
(\dim P_{i_1}\ldots\dim P_{i_{n-1}})^2}\right]
 \log  
\left[{
P_{i_{n-1}}\ldots P_{i_1}\rho P_{i_1}\ldots P_{i_{n-1}}
\over 
(\dim P_{i_1}\ldots\dim P_{i_{n-1}})^2}\right]
\right) \nonumber\\
& &\qquad -2\log\dim{\cal V}\nonumber\\
&=&
-\sum_{i_1,i_2,\ldots,i_{n-1}} 
 \mbox{tr}\left(\left[{P_{i_{n-1}}\ldots P_{i_1}\rho P_{i_1}\ldots P_{i_{n-1}}
}\right] \log\left[{P_{i_{n-1}}\ldots P_{i_1}\rho P_{i_1}\ldots P_{i_{n-1}}
\over (\dim P_{i_1}\ldots\dim P_{i_{n-1}})^2}\right]\right)\nonumber\\
& &\qquad-2\log\dim{\cal V}. \label{rhs1} 
\end{eqnarray}
We may now use (\ref{KlogK2}) to give
\begin{eqnarray}
\lefteqn{I_{d,W}} 											\nonumber\\
&\geq& -\sum_{i_1,i_2,\ldots,i_{n-1}}
 \mbox{tr}\left(\left[{P_{i_{n-1}}\ldots P_{i_1}\rho P_{i_1}\ldots P_{i_{n-1}}
}\right] \log 
\left[{P_{i_{n-1}}\ldots P_{i_1}\rho P_{i_1}\ldots P_{i_{n-1}}
\over
	(\dim P_{i_1}\ldots\dim P_{i_{n-1}})^2}\right]\right)-2\log\dim{\cal V}	\nonumber\\
 &=& 
-\sum_{i_1,i_2,\ldots,i_{n-2}}(\dim P_{i_1}\ldots\dim P_{i_{n-2}})^2 \nonumber\\
	&& \quad\times\sum_{i_{n-1}} \mbox{tr}\Big(P_{i_{n-1}}
\left[{P_{i_{n-2}}\ldots P_{i_1}\rho P_{i_1}\ldots P_{i_{n-2}}
\over 
(\dim P_{i_1}\ldots\dim P_{i_{n-2}})^2}\right]P_{i_{n-1}}	\nonumber\\ 
	& &\qquad\times \log {P_{i_{n-1}}\over (\dim P_{i_{n-1}})^2}
		\left[{P_{i_{n-2}}\ldots P_{i_1}\rho P_{i_1}\ldots P_{i_{n-2}}
\over 
(\dim P_{i_1}\ldots\dim P_{i_{n-2}})^2}\right]P_{i_{n-1}}
\Big)-2\log\dim{\cal V} 														\nonumber\\
	&\geq& -\sum_{i_1,i_2,\ldots,i_{n-2}} 
 		\mbox{tr}\left(
\left[{P_{i_{n-2}}\ldots P_{i_1}\rho P_{i_1}\ldots P_{i_{n-2}}}\right]
 \log\left[{P_{i_{n-2}}\ldots P_{i_1}\rho P_{i_1}\ldots P_{i_{n-2}}
\over 
(\dim P_{i_1}\ldots\dim P_{i_{n-2}})^2}\right]\right) -2\log\dim{\cal V}.
\nonumber\\
& &
\end{eqnarray}

The right-hand-side of this expression is of the same form as the
right-hand-side of (\ref{rhs1}) but with one less summation ($n-2$ summations
compared to $n-1$ in (\ref{rhs1})).  We may now repeat this step 
recursively to show that
\begin{equation}
I_{d,W} \geq  \mbox{tr}\left(\rho\log\rho\right)-2\log\dim{\cal V}.
\end{equation}

	Thus the minimum of $I_{d,W}$ over consistent windows containing
only homogeneous histories is $-\mbox{tr}(\rho\log\rho)-2\log\dim{\cal V}$.  In other
words, all the information-entropy lies in the initial state for
standard $n$ time quantum theory with unitary evolution.

	It is worth noting that if the time evolution is {\em
non\/}-unitary (such as might occur in a space-time region around a
black hole) then histories such as (\ref{Uproj}) cannot be used to
minimise the information-entropy since the operators at each time
are no longer projectors.  Thus there will be a contribution to the
information-entropy from the time evolution in addition to that from
the initial state.

\subsection{Two-time histories}
As was remarked earlier, we suspect that $-\mbox{tr}(\rho\log\rho)-2\log\dim{\cal V}$
is the minimum of $I_{d,W}$ over all consistent sets although, so
far, we have only been able to show this in certain special cases.
One such is the two-time history version of a quantum system with canonical
Hilbert space ${\cal H} = {\Bbb C}^N$ ({\em i.e.,\ } the history Hilbert space is ${\cal V}
= {\cal H}\otimes {\cal H} = {\Bbb C}^{N^2}$) with a unitary time evolution and
the special initial density matrix   
\begin{equation}
\rho = {\rm diag}({1\over N},{1\over N},{1\over N},\ldots,{1\over N}). 
											\label{rho_diag}
\end{equation}
In fact we may take the time evolution to be trivial ({\em i.e.} we 
choose the Hamiltonian to be zero) without loss of generality, since 
the class of history propositions we will consider takes into account 
all possible unitary evolutions.

We now show that the minimum of $I_{d,W}$ over {\em all\/}
consistent sets is
\begin{equation}
	-\mbox{tr}_{\cal H}(\rho\log\rho)-2\log\dim{\cal V} = \log N-2\log N^2
=-3\log N.
\end{equation}

	The most general unit vector in the tensor product space ${\cal V} =
{\cal H}\otimes{\cal H}$ is
\begin{equation}
	v = \sum_{i,j=1}^N v^{ij}  |i\rangle\otimes |j\rangle \label{vector}
\end{equation}
where $\{ |i\rangle \}_{i=1}^N$ is an orthonormal basis for ${\cal H}$ and
the constants $ v^{ij}$ satisfy
\begin{equation}
	\sum_{ij} v^{ij}v^{*ij} = 1.
\end{equation}
The one-dimensional projection operator onto the subspace defined by
this vector is
\begin{eqnarray}
	P_v &=& \sum_{ijkm}\left( v^{ij} |i\rangle\otimes|j\rangle\right)\left(
		v^{*km}\langle{k}| \otimes\langle{m}|\right)			\nonumber\\ 
		&=& \sum_{ijkm}
	v^{ij}v^{*km}\ |i\rangle\langle{k}|\otimes|j\rangle\langle{m}|.
\label{proj_v} \end{eqnarray}

	If we consider consistent sets that contain only one-dimensional
projectors then in order to decrease the information-entropy at
least one projector must have a probability greater than $1/N$.
However, we will now show that the maximum value of the probability
of any one-dimensional projector in a consistent set is $1/N$,
so that no consistent windows with one-dimensional histories reduces
the information-entropy below $\log N-2\log N^2 = -3\log N$, the value obtained by
considering windows with only homogeneous histories.  We also show
that (as might be expected) windows with higher-dimensional
histories also fail to reduce the information-entropy below this
value.

	If $P_v$ is part of a consistent set then
\begin{equation}
	d(1-P_v,P_v)= 0
\end{equation}
so that
\begin{equation}
d(1,P_v) = d(P_v + (1-P_v),P_v) = d(P_v,P_v) + d(1-P_v,P_v) = d(P_v,P_v)
\end{equation}
where we have used the fact that if $\alpha$ and $\beta$ are two disjoint
histories, then for any other history $\gamma$,
\begin{equation}
	d(\alpha\oplus\beta,\gamma)=d(\alpha,\gamma) + d(\beta,\gamma).
\end{equation}
Thus the probability of this particular consistent history
proposition is 
\begin{eqnarray}
 d(P_v,P_v)&=&d(1,P_v)									\nonumber\\
&=&\mbox{tr}_{{\cal V}\otimes{\cal V}}([1\otimes P_v]\ X)
\end{eqnarray}
where $X$ is the decoherence operator for this system which may be found
in \cite{IL94b} (as a special case of the results given there):
\begin{equation}
X= [R_{(2)} \otimes 1_2]S_4 [1_2\otimes (\rho\otimes 1_1)][R_{(2)} \otimes 1_2].
\end{equation}
In this expression, $1_1$ is the unit operator on ${\cal H}$ and $1_2$ is 
the unit operator on ${\cal V} = {\cal H}\otimes{\cal H}$; $R_{(2)}$ is the
`time-reversal' operator on ${\cal V} = {\cal H}\otimes{\cal H}$:
\begin{equation}
R_{(2)} u_1\otimes u_2 = u_2\otimes u_1;\quad R_{(2)}^2 = 1;
\end{equation}
and $S_4$ is the map on $\otimes^4 {\cal H}$ which acts as
\begin{equation}
S_4 (u_1\otimes u_2\otimes u_3\otimes u_4 )=  u_2\otimes u_3\otimes u_4\otimes u_1,
\end{equation}
which has the important property that, for any four operators $A,B,C,D$ on ${\cal H}$,
\begin{equation}
\mbox{tr}_{\otimes^4{\cal H}}([A\otimes B\otimes C\otimes D]\ S_4) = 
\mbox{tr}_{{\cal H}}(A  B  C  D )\label{S4}.
\end{equation}
Thus,
\begin{eqnarray}
 d(1,P_v)&=& \mbox{tr}_{{\cal V}\otimes{\cal V}}([1_2\otimes P_v]\ X)\nonumber\\
&=& \mbox{tr}_{\otimes^4{\cal H}}([1_2\otimes P_v][R_{(2)} \otimes 1_2]S_4 
         [1_2\otimes (\rho\otimes 1_1)][R_{(2)} \otimes 1_2])\nonumber\\
&=& \mbox{tr}_{\otimes^4{\cal H}}([R_{(2)} \otimes 1_2][1_2\otimes P_v][R_{(2)} \otimes
1_2]S_4 
         [1_2\otimes (\rho\otimes 1_1)])\nonumber\\
&=& \mbox{tr}_{\otimes^4{\cal H}}([1_2\otimes P_v]S_4 
         [1_2\otimes (\rho\otimes 1_1)])\nonumber\\
&=& \mbox{tr}_{\otimes^4{\cal H}}(S_4 
         [1_2\otimes \{(\rho\otimes 1_1)P_v\}])\nonumber\\
&=&\sum_{ijkm}v^{ij}v^{*km}\mbox{tr}_{\otimes^4{\cal H}}(S_4 [1_1\otimes 1_1 \otimes \rho
|i\rangle\langle{k}|\otimes |j\rangle\langle{m}|])\nonumber\\
&=&\sum_{ijkm}v^{ij}v^{*km}\mbox{tr}_{\cal
H}[\rho|i\rangle\langle{k}|j\rangle\langle{m}|] 
\end{eqnarray}
where we have used (\ref{S4}).  

Thus
\begin{eqnarray}
d(1,P_v)&=&{1\over N}
\sum_{ijkm}v^{ij}v^{*km}\delta^{im}\delta^{kj} \nonumber\\ 
&=&{1\over N}
\sum_{ij}v^{ij}v^{*ji}					\nonumber\\ 
&=&{1\over N} \left(
\sum_{i}v^{ii}v^{*ii} + \sum_{i\neq j}v^{ij}v^{*ji}
    \right)												\nonumber\\ 
&=&{1\over N} \left( \sum_{i}v^{ii}v^{*ii} + \sum_{i<
j}\left(-|v^{ij}-v^{ji}|^2 + v^{ij}v^{*ij}
    +v^{ji}v^{*ji}\right)
    \right)												\nonumber\\ 
&=&{1\over N} \left( 1-  \sum_{i< j}|v^{ij}-v^{ji}|^2
    \right)												\nonumber\\
&\leq& {1\over N}.
\end{eqnarray}
We also note that this calculation shows that for a $k$ dimensional
projector $P_{(k)}$, the probability
\begin{equation}
	d(P_{(k)},P_{(k)})=d(P_{(k)},1) \leq {k\over N}
\end{equation}
so that including higher dimensional projectors in a window will
only increase the value of the information-entropy.  Hence, for this
example, $-\mbox{tr}_{\cal H}(\rho\log\rho) -2\log\dim{\cal V}= -3\log N$ is the minimum value of
$I_{d,W}$ over all windows.

\section{CONCLUSION}
We have shown that there are no pure decoherence functions in the
consistent histories approach to generalised quantum theory since
every decoherence function may be written as the sum of two others.

More substantially, we have also put forward a definition of
information-entropy in generalised quantum mechanics that relies crucially on the notion of the
dimension of a history, a concept that is natural within our
approach to the general scheme.  It is worth noting that fundamental
to the consistent histories approach from the start has been the
idea of taking the sum of two homogeneous histories in standard
$n$-time quantum theory to form inhomogeneous histories.
However, the idea of the dimension of an inhomogeneous history is
difficult to understand unless, as we have frequently advocated,
histories are identified with projection operators on an $n$-fold
tensor product space.

We have called the function $I_{d,W}$ a measure of information-entropy for 
generalised quantum mechanics as it has key properties that it decreases under
refinement and it is small for consistent windows in which the probability is
peaked around histories of small dimension (as we have shown, 
decoherence functions may or may not have such windows).  The fact that $I_{d,W}$ is negative,
however means that it is not quite a usual measure of missing information. One
does, of course, have the option of using the negative of the function
$I_{d,W}$, however 
we have not done so in order to facilitate comparison with other approaches.

It ought to be said at this stage that while the function $I_{d,W}$ 
has many of the properties that one requires of a measure of 
information-entropy in the
space-time context, its true meaning is still somewhat unclear.  In this context it should be noted that any function of the form
\begin{equation}
I_{d,W}^x = -\sum_{i=1}^n d(\alpha_i,\alpha_i)
\log \left[{d(\alpha_i,\alpha_i)
\over \left(\dim\alpha_i/\dim{\cal V}\right)^x}\right]\end{equation}
where 	$x\geq 1$ is a real number also has the key property that it 
decreases under refinement of the consistent set.  The case $x=1$ 
may turn out to be the most interesting, as in this case, the 
measure of information is (minus) the 
Kullback information \cite{Kull59} of the 
distribution $\{d(\alpha_i,\alpha_i)\}$ relative to 
a `maximally ignorant' distribution on the set $\{\alpha_i\}$ which has 
${\rm Prob}(\alpha_i) = {\rm dim}(\alpha_i)/{\rm dim}{\cal V}$.  The relationship between the measures with different
values of $x$ needs to be understood.
Interestingly,
Gell-Mann and Hartle \cite{GH95} have considered measures of this sort as a
result of rather different considerations such as thermodynamic 
depth \cite{LloydPagels88}.  We understand \cite{private} that they 
have also considered a measure of entropy 
which they call a \lq bundle of histories entropy\rq\ which 
takes into account the number of fine-grained histories in a coarse-grained 
history; this idea is clearly related to the one we have put forward.

We anticipate that our definition of information-entropy---which
is a straightforward function on the class of consistent sets with
attractive properties under refinement---may help in the development
of a set selection criterion\footnote{The importance of this issue
for the whole framework has been discussed by Dowker and Kent\cite{DK96}.}: for example, in the case that
the system naturally divides into a subsystem and the \lq environment\rq, this might be done 
 by selecting the set which
minimises the information-entropy of the distinguished subsystem (see for example \cite{Zurek94}).    This is an important problem to which we
intend to return in future work. Related issues that need to be
understood are the r\^ole of symmetries (see for example
\cite{Schr96a,Schr96b,BrunHall95}) and the existence of
quasi-classical domains and their relation to the system-environment
split.  In the context of the latter, it should be noted that if our
vector space ${\cal V}$ happens to arise as the tensor product of two
spaces ${\cal V}_1$ and ${\cal V}_2$, then our definition of information-entropy
has precisely the behaviour that might be hoped for.  For if one
considers a consistent window in which each history proposition $\alpha$
is a tensor product $\alpha =\alpha_1\otimes\alpha_2$, with $\alpha_1 \in P({\cal V}_1)$
and $\alpha_2 \in P({\cal V}_2)$, then the information-entropy is the sum of
the information-entropy associated to each sub-system.

\bigskip 
\section*{Acknowledgements}

\noindent
We are very grateful to Jim Hartle and Adrian Kent for reading an earlier 
draft of this paper and for their many penetrating comments.
We would also like to thank Jeremy Butterfield and Sandu Popescu
for many helpful discussions. We are very grateful to the
Leverhulme and Newton Trusts for the financial support given to one
of us (NL).

 \end{document}